\documentclass{acm_proc_article-sp}

\newcommand{\aiu}{\texttt{AIU }}
\newcommand{\aius}{\texttt{AIUs }}

\newcommand {\à} {\`{a}}

\begin{document}

\title{Modelling internet based applications for designing multi-device adaptive interfaces}
%
%

\numberofauthors{2}
%

\author{
%
\alignauthor Enrico Bertini and Giuseppe Santucci\\
        \vspace{0.2cm}
       \affaddr{Dipartimento di Informatica e Sistemistica
        Universit\à di Roma "La Sapienza"\\
       \affaddr{Via Salaria, 113}}\\
       \affaddr{Roma, Italy}\\
       \email{\{bertini, santucci\}@dis.uniroma1.it}
}
\maketitle
\begin{abstract}
\thispagestyle{empty} \noindent

 The wide spread of mobile devices in the consumer market has
  posed a number of new issues in the design of internet applications and their user
  interfaces.  In particular, applications need to adapt their interaction modalities to different
  portable devices.
In this paper we address the problem of defining models and techniques for
  designing internet based applications that automatically adapt to different mobile devices.  First, we define a formal model that allows for specifying  the interaction  in a way that is abstract enough to be decoupled from the presentation layer,
  which is to be adapted to different contexts.  The model is mainly based on the idea of describing the user interaction in terms of elementary actions. Then, we provide a formal device characterization showing how to effectively implements the \aius in a multidevice context.

\end{abstract}

\vspace{-2mm}\section{Introduction}
The era of standard situated PCs is over; nowadays users own
multiple different computing appliances that can be easily
transported and that, consequently, operate in multiple changing
environments. In order to address this issue
the next generation of internet based application should be designed for adapting
to a very large spectrum of different characteristics. This is why the research is faced with the
problem of finding models and techniques to design applications
aware of and adaptable to: (a) Devices (e.g., cellphones, PDAs, PCs), (b) Environment (e.g., noisy room, low light, moving person), and Users (e.g., users with special needs).\\
In this paper we focus on the first issue trying to solve different problems: architectures that support heterogeneous devices, models
to exchange and share data,  techniques for building effective
user interfaces. Our focus is on user interfaces and our current
approach is mainly based on the idea of modelling user interaction
in very abstract and simple way. 
Moreover, we would like to clarify the context in which our
proposal takes place. We want to deal with simple internet based
applications/services, like reserving an hotel room, finding a
restaurant, or  booking a flight seat. It is out of the scope of
our approach to redesign generic web sites making them  accessible
through different devices or implementing complex applications
over large information systems. We are focusing on the plethora of
simple but useful applications that can run on a cellular phone,
on a connected PDA, and so on, providing the user with concise
pieces of information and/or simple services.\\
The rest of the document is structured as follows: in
Section~\ref{sec:related-work} we describe some related proposals, in
Section~\ref{sec:aius} we introduce a
model that supports the design of adaptive applications: the \emph{Atomic Interaction Units} model and a
working example is provided. Section~\ref{sec:devices},
provides a description of the main device characteristics we want to take into account.
Section~\ref{sec:implementation} describes some preliminary results on the \aius implementation strategies, and finally in Section~\ref{sec:conclusions} some conclusions and open issues are discussed.

\vspace{-2mm}\section{Related works}
\label{sec:related-work}

The problem of generating different interfaces for different
devices  is often indicated as
the problem of creating \emph{plastic interfaces}
\cite{plastic-ui}.\\
Many ideas come from  research in \emph{model-based} user
interface design \cite{model-based-uis} where the designer is
supposed to design an interactive system by editing and
manipulating abstract models (e.g., task model) that describe the
system's behavior and where the system is supposed to
automatically generate the final application code.\\
In \cite{vanderdonckt-encapsulation, puerta-mobile, puerta-adaptation}  a design 
framework that consists of a \emph{platform model}, a
\emph{presentation model}, and a \emph{task model} is presented. The designer
builds the user interface by means of abstract interaction objects
(AIOs) that are platform-neutral widgets he can assemble to design
the interface in an abstract manner. Different presentation
structures can then be generated to allocate the interaction units
across many windows.\\
With a similar approach, in \cite{paterno-nomadic} the Teresa
tools is presented. It provides an environment to design
multi-device applications that is strongly based on task
modelling and the
presentation structure is automatically generated from a task
model, taking into account temporal constraints specified into it.
Screen space optimization is obtained mapping the abstract objects
into suitable concrete objects.\\
The same idea of heavily exploiting formal models to design
interactive applications comes from research on data-intensive web
design, as illustrated in \cite{data-intensive-survey}, that stems
from past research on model-based hypermedia design, like RMM
\cite{rmm} and HDM \cite{hdm}, and that has a major focus on data
modelling. This approach suggests a process in which the designer
starts from a model of the data (usually drawing an
entity-relationship model) and on top of it creates an hypertext
model that defines the connection points between the web pages,
i.e., the links. Finally, he creates a presentation part that
permits to give a visual form to the various pages. WebML
\cite{webml} is a powerful data-driven language that permits to
describe an hypertext composed of single atomic blocks that are
tightly connected to underlying data elements.\\
Our work shows similarities with all these systems, in fact here
we propose a model-based approach. We adopt the idea of
abstracting on interaction elements and provide a collection of
atomic interaction units that are the abstract counterpart of
common interaction elements. At the same time our work diverges
from this approach and comes closer to data-intensive web
modelling approach. Hence, we propose to directly design the
hypertext and, thus, the structure of links that connect the
presentation units. As it will be described in detail below, our
method fundamentally consists in specifying a graph structure (an
UML Activity Diagram) in which the nodes are populated with atomic
interaction units and the edges represent the transition triggered
interacting with atomic interaction units.

\vspace{-2mm}\section{The Atomic Interaction Units}
\label{sec:aius}
The foundation of our proposal is
an abstract model able to describe the user interaction with the system focusing on the basic activities whose composition will produce simple but effective internet based application. As a consequence, we model the information that is exchanged between the user and the system together with the purpose for which such an information is exchanged.
Using this approach the designer is provided with a
formalism to specify the information content of each presentation and the
connection among the various parts, in order to indicate the
behavior of the application, that is, how the system evolves as
the user interact with it.\\
Our proposal consists of two main
parts:(a) a set of Abstract Interaction Units (\aius) to be used as
building blocks for abstract interface definition and (b)
the UML \emph{Activity Diagram} as formalism to connect the  \aius that compose the interface.
The set of \aius has been designed analyzing the user interfaces
that are actually used to model standard web services. The challenge is in collecting a small set
of atomic units that could describe the interaction, abstract enough
to be completely unrelated with a particular device  but expressive enough to let designers to 
model complex services. The effort we made has produced a small
set of \aius, described in the following section.\\
The UML \emph{Activity Diagram} is basically a state chart diagram
in which each state represents an activity and each transition is
triggered by the end of this activity. How the \emph{Activity Diagram} can
be used to glue together the \aius will be explained later by using
an example.\\
\begin{figure*}
\centering \resizebox{16cm}{!}{\includegraphics{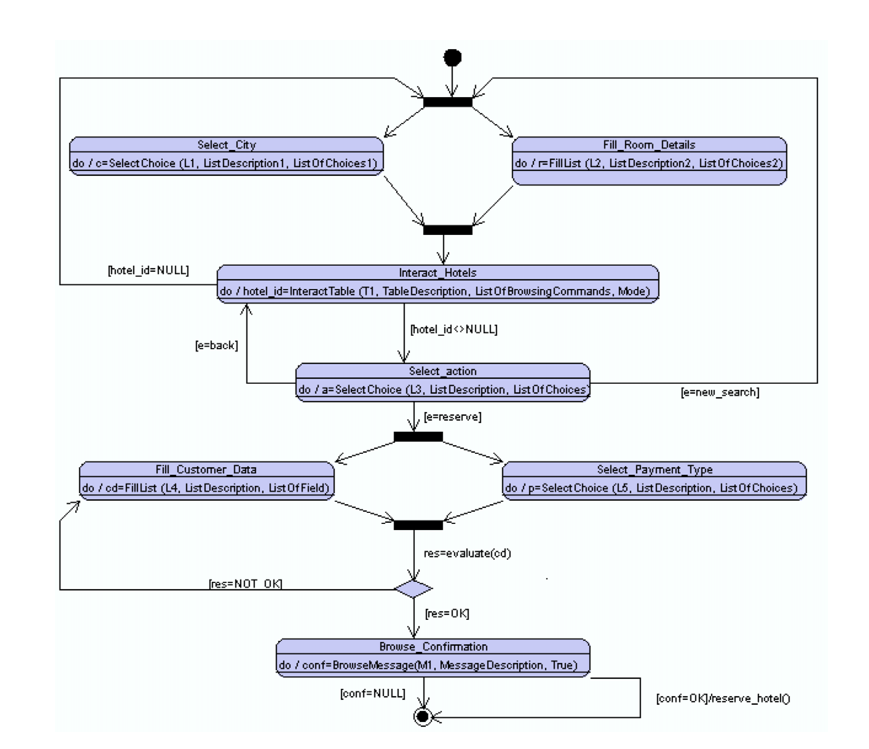}}
\caption{{\em Activity diagram and \aius to model a service that
permits an  hotel reservation}\label{fig:aiu-example}}
\vspace{-0mm}
\end{figure*}
We foresee two main interaction activities: browsing, i.e., just
observing something produced by the system and inputting, i.e.,
providing the system with some information. 
In the following each \aiu is described in detail. 
Note that all the \aius share a Quit command that allows for leaving the 
\aiu with no effects and returning the null value.

\vspace{-5mm}   
\begin{verbatim}
BrowseImage(ImageId,ImageDescription,BrowsingCommands)
:{NULL, elemOfBrowsingCommands}
\end{verbatim}
\vspace{-5mm}   

This \aiu allows for browsing an image; usual
facilities of zooming and panning are provided, if possible, by
the device. The image description is a two values record:
[\texttt{ImageName}, \texttt{ImageSummary}], where
\texttt{ImageName} is used as a title during the image
presentation and \texttt{ImageSummary} is an image description
that can be used when the video channel is not available or
disturbed. The \texttt{BrowsingCommands} is a set of
commands oriented towards server side image manipulation (e.g.,
changing image detail and/or resolution); such commands do not
allow to reach any other state than the one hosting the \aiu
(i.e., they correspond to self-transitions).

\vspace{-5mm}
\begin{verbatim}
InteractImage(ImageId,ImageDescription,BrowsingCommands)
:{point, NULL, elemOfBrowsingCommands}
\end{verbatim}
\vspace{-5mm}      

This \aiu is quite similar to the \texttt{BrowseImage}; the only difference is that
the user can leave the \aiu both choosing the Quit button or selecting a point
(the \aiu returns the x,y coordinates of a point) on the image itself.
    
\vspace{-5mm}
\begin{verbatim}
BrowseText(TextId,TextDescription,BrowsingCommands)
:{NULL, elemOfBrowsingCommands}
\end{verbatim}
\vspace{-5mm}

This \aiu allows for browsing a large body of
text. The text description is a two value record:
[\texttt{TextName}, \texttt{TextSummary}], where \texttt{TextName}
is used as a title during the text presentation and
\texttt{TextSummary} is a text description that can be used when
the video channel is not available or disturbed or as an
alternative when the device capability of displaying such an
amount of text is very poor.

\vspace{-5mm}
\begin{verbatim}
BrowseMessage(MessageId,MessageDescription, OKbutton)
:{NULL, OK}
\end{verbatim}
\vspace{-5mm}

This \aiu allows for browsing a message;
usual facilities of panning are provided, if possible, by the
device. \texttt{OKbutton}
is a boolean parameter that forces the activation of an OK
command, useful to answer messages that contain a confirmation
choice.

\vspace{-5mm}
\begin{verbatim}
BrowseTable(TableId, TableDescription,BrowsingCommands)
:{NULL, elemOfBrowsingCommands}
\end{verbatim}
\vspace{-5mm}

This \aiu allows for browsing a relational
table; usual facilities of panning are provided, if possible, by
the device.

\vspace{-5mm}
\begin{verbatim}
InteractTable(TableId,TableDescription,BrowsingCommands)
:{NULL, elemOfBrowsingCommands,tableTuple)
\end{verbatim}
\vspace{-5mm}

This \aiu is quite similar to the
\texttt{BrowseTable} \aiu. The main difference is that the user
can leave the \aiu by selecting a tuple.

\vspace{-5mm}
\begin{verbatim}
FillList (ListId,ListDescription,ListOfField)
:{NULL, ListOfField}
\end{verbatim}
\vspace{-5mm}

The \texttt{FillList} \aiu allows for filling a list of field. 
The \aiu returns the list with the filled
values.

\vspace{-5mm}
\begin{verbatim}
SelectChoice (ListId,ListDescription,ListOfChoices)
:{NULL, elementOfListOfListOfChoices}
\end{verbatim}
\vspace{-5mm}

This \aiu allows for selecting an element among a
predefined  a list of Choice and returns the selected element.

\vspace{-5mm}
\begin{verbatim}
SelectMutipleChoice(ListId,ListDescription,ListOfChoices)
:{NULL, elementOfListOfChoices}
\end{verbatim}
\vspace{-5mm}

This \aiu allows for selecting one or more
elements among a predefined  list of choices, returning the selected
values.

\vspace{-2mm}\subsection{AIUs at work}
\label{sec:aiuexample}
In this section we want to
describe how the composition of these units can lead to the design
of a whole service. The UML Activity Diagram is used to compose
the \aius and define the service. Each activity state contains
one \aiu. Transitions are triggered by the user acting with the interaction units
and each transition correspond to a computation operated on the
server. Some interaction units can also appear in parallel by
using the fork construct. This takes into account the common
situation in which a single presentation contain more than one
\aiu at the same time and the case in which we are modelling a task
that involves interactions that do not have a pre-defined
sequential ordering.  In order to clarify the use of this model here we provide an
example describing an hotel reservation service (see Figure~\ref{fig:aiu-example}).
The user starts inputting data about the hotel city and some details about the period he wants to reserve.
Since these are two separate tasks they are modelled with two
separate \aius. The city specification is a \emph{SelectChoice}
\aiu, the details specification is a \emph{FillList} \aiu for
which the user is requested to input data about the reservation
period. The order of these two task is irrelevant, so they are
connected with a fork construct. The final implementation could be
a whatever ordering of these tasks or, if the selected device has
enough screen space, a single unified view of these two tasks.
As the the user sends input data, the system passes to the next
activity. The result
is modelled as an \emph{InteractTable} \aiu. The result of the
query (search for an hotel), in fact, is a set of objects characterized by set of attributes. 
When the user selects a certain hotel, the system moves to the
next activity; the selection of an action to perform on it. The
transition between the "Interact\_Hotels" activity to the
"Select\_Action" activity involves a parameter passing. The
\emph{InteractTable} \aiu, as described before, has a parameter in
output the system sends as the user selects an object from the
table. In the "Select\_Action" activity, the user is requested for
selecting an action to perform between the following list: reserve
the hotel, start a new search, return back to the result. This is
modelled with a \emph{SelectChoice} \aiu.
According to the selection, the system can proceed to three
different activities: 1)return to the starting point, 2)go back to the previous result, and 3)proceed with the reservation task. 
In the last case the system steps forward to a new fork hosting
two concurrent activities: the customer data
specification and the selection of the payment method. The
"Fill\_Customer\_Data" activity is modelled as a \emph{FillList}
\aiu because it is supposed to accept data directly specified by
the user. The "Select\_Payment\_Type" activity is modelled as a
\emph{SelectChoice} \aiu because it is supposed to present the user with a
predefined list of payment methods.
Eventually the system checks the data and, in case of error, redirects the user to the form. 
Otherwise the system
requests for a confirmation and, when the user confirms, the
system collects the data and makes the reservation.
This  example shows how the composition of the abstract
interaction units can be done in order to model a service. After
this phase, the system must be able to translate this model into a
final implementation. In order to perform the translation the
system must take into account a set of key device characteristics, described in the next section.

\vspace{-2mm}\section{Devices and {AIUs} metrics}
\label{sec:devices}

In order to effectively implement the \aius on physical devices,
we need some figures about their
capabilities. Different classifications and characteristics are
available in the literature (e.g., \cite{device-categories}).
Here, we focus on a first set of characteristics that constitute
the minimal information needed to adapt the different \aius to
each device. Moreover, we need to investigate the \aius as well
because of the size of the parameters they handle heavily affects
their implementation (e.g., the way in which the user interacts
with a relational table may differs depending on the number of
tuples and attributes). A practical usage of such parameters is shown in
Section~\ref{sec:implementation}. Finally, in this work, even if
we devised some \aius oriented towards image manipulation, we
concentrate on textual based \aius and, consequently, we consider only text/table oriented metrics.
Concerning \textbf{devices} we define the following functions:
\begin{itemize}
\vspace{-2mm}
\item int RN(dev) (Row Number), returning the number of rows the
device is able to display; 
\vspace{-2mm}
\item int CN(dev) (Column Number),
returning the number of columns the device is able to display;
\vspace{-2mm}
\item boolean CVS(dev) (Continuous Vertical Scrolling), returning
the availability of a continuous (i.e., pixel based)  vertical
scrolling; 
\vspace{-2mm}
\item boolean RVS(dev) (Row-based Vertical Scrolling),
returning the availability of stepped (i.e., row based)  vertical
scrolling; 
\vspace{-2mm}
\item boolean PVS(dev) (Page-based Vertical Scrolling),
returning the availability of stepped (i.e., page based) vertical
scrolling; 
\vspace{-2mm}
\item boolean CNHS(dev) (CoNtinuous Horizontal
Scrolling), returning the availability of a continuous (i.e.,
pixel based)  horizontal scrolling; 
\vspace{-2mm}
\item boolean COHS(dev)
(COlumn-based Horizontal Scrolling), returning the availability of
stepped (i.e., column based)  horizontal scrolling; 
\vspace{-2mm}
\item boolean
PHS(dev) (Page-based Horizontal Scrolling), returning the
availability of stepped (i.e., page based) horizontal scrolling;
\item boolean WE(dev) (WAP Enabled), true if the device is Wap
enabled; 
\vspace{-2mm}
\item boolean JE(dev) (Java Enabled), true if the device
is Java enabled; 
\vspace{-2mm}
\item boolean AA(dev) (Audio Availability),
returning the audio channel availability; 
\vspace{-2mm}
\item int CD(dev) (Color
Depth), returning the color/black and withe depth (expressed in
bit); 
\vspace{-2mm}
\item boolean TSA(dev) (Touchable Screen Availability),
returning the availability of touchable surfaces.
\end{itemize}

\begin{figure*}
\centering \resizebox{18cm}{!}{\includegraphics{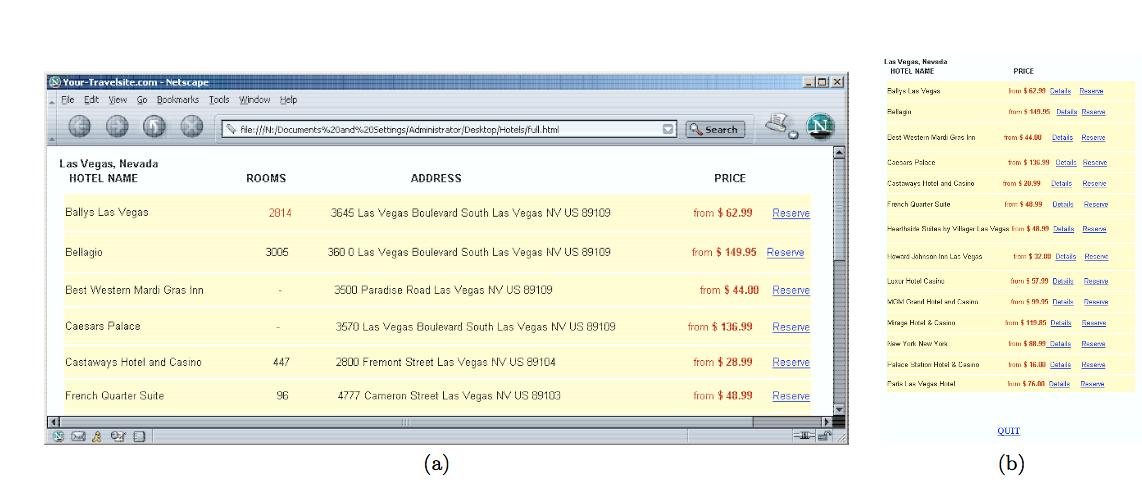}}
\caption{{\em Implementing the InteractionTable \aiu on a
regular browser (a) or on a
medium size handheld (b)}} 
\label{fig:table} \vspace{-0mm}
\end{figure*}


Concerning  \textbf{\aius}, we investigate some metrics about the text based \aius, distinguishing between table and pure text oriented \aius.

\begin{itemize}
\vspace{-2mm}\item int RN(table-oriented-\aiu), (Row Number) returning the number of rows the \aiu needs to be displayed;
\vspace{-2mm}\item int CN(table-oriented-\aiu), (Column Number) returning the number of columns the \aiu needs to be displayed;
\vspace{-2mm}\item int CHN(text-oriented-\aiu), (CHaracter Number) returning the number of characters the \aiu needs to be displayed.
\end{itemize}

The main idea, described in the next session through an example,
is to use the above metrics to measure the level of degradation an
\aiu suffers when implemented on a specific device. As an example,
we can count the number of scrolling commands (both vertical and
horizontal) needed in order to display the overall \aiu content
and, depending on the computed figures, decide to rearrange the
\aiu structure, adapting it to the particular device.

\vspace{-2mm}\section{Implementation issues}
\label{sec:implementation}
In this section we propose a
dynamic adaptation based on the metrics described in the previous section.
A systematic analysis of the \aius implementation is out of the
scope of this paper. Here we discuss, as a working example, the
implementation on the device $d$ of the \aiu $a$,  InteractTable
described in the example shown in Section~\ref{sec:aiuexample}.
The purpose of such an \aiu is to allow the user for displaying a
table containing a set of hotels (one per row) allowing the
selection of a specific hotel. Assume that the following figures
hold for the involved device and \aiu:

\begin{itemize}
\vspace{-2mm}\item RN($a$)= 40 (i.e., the tables contains 40 hotels);
\vspace{-2mm}\item CN($a$)= 105 (i.e., each row needs 105 columns);
\vspace{-2mm}\item CN($d$)= 30 (i.e., the device can handle 30 columns);
\vspace{-2mm}\item RN($d$)= 14 (i.e., the device can handle 14 rows);
\vspace{-2mm}\item RVS($d$)= true (i.e., the device allows for row based vertical scrolling);
\vspace{-2mm}\item PVS($d$)= true (i.e., the device allows for page based vertical scrolling);
\vspace{-2mm}\item COHS($d$)= false (i.e., the device does not allow for column based horizontal scrolling);
\vspace{-2mm}\item PHS($d$)= true (i.e., the device does not allow for page based horizontal scrolling);
\vspace{-2mm}\item JE($d$)=false (i.e., the device is not Java enabled);
\vspace{-2mm}\item WE($d$)=true (i.e., the device is Wap enabled);
\end{itemize}

Based on these figures, we can argue that the \aiu can be easily
displayed for what concerns the number of columns: in fact, it
requires at most 40 row based scrolling commands or $\lceil 40/14
\rceil$ page based vertical scrolling commands and such figures
are quite reasonable (formally speaking, we define some threshold
values). On the other hand, handling 105 columns on a device that
is able to display only 30 columns and does not allows for
horizontal scrolling is not an easy task. The only way is to
present the table to the user in a two steps interaction: (1) the user is presented with a table containing only a subset of the table attributes (e.g., hotel-name and hotel-price) whose column occupation is less
 than 30 and (2) and additional commands allows for detailing a single
row, getting all the hotel attributes.
The way in which it is possible to implement the above strategy
strongly depends on the device computational capabilities (e.g.,
Java enabled) and on load balancing issues. Here, in order to
follow the more robust solution we assume that all the work is
performed by the server that, looking at the device capabilities,
will produce the needed Java code or Wap pages. As an example, we
can see in Figure~\ref{fig:table} (a) a possible implementation of
the InteractTable \aiu on a device with no significative
limitations (e.g, a usual browser on a portable PC).
In order to display the same table on a device with the above
capabilities we have to reduce the table attributes, producing the
table shown in Figure~\ref{fig:table} (b). In such a table only
the hotel-name and the hotel-price are available; if the user
requires more pieces of information, a new command, details, is
available showing the full hotel description.\\
The above example provides the feeling on how it is possible to
adapt the same \aiu to different device. We are currently
investigating different implementation strategies and different
threshold values to in order to come up with a more formal and
complete way of implementing \aius.

\vspace{-2mm}\section{Future work and conclusions}
\label{sec:conclusions}

In this paper we presented a novel approach for implementing, on a
variety of portable devices, simple internet based applications.
The main ideas supporting our proposal are (a) a formal model characterizing the user interaction building blocks (\aius), (b) the embedding of such a model within the UML activity diagram, (c) the formalization of the characteristics of devices and \aius  through several suitable metrics, and
(d) the definition of ad-hoc strategies for implementing in efficient way the \aius on different devices.
Some aspects of our approach deserve more deep analysis: we are
currently working on defining a complete set of \aius and devices
metrics; moreover we are developing a first prototype for
implementing the \aius, in order to have some feedback about our
approach.\\
\textbf{Acknowledgments}
Work supported by the MIUR-FIRB project "MAIS" (Multichannel
adaptive Information Systems, http://black.elet.polimi.it/mais/index.php).
\vspace{-2mm}\bibliographystyle{abbrv}
\bibliography{main-bib}
\end{document}